\documentclass{elsart}
\usepackage{graphics}
\usepackage{amssymb}
\usepackage{epsfig}
\begin{document}

\begin{frontmatter}

\title{Q-ball Formation in Affleck-Dine Baryogenesis with
Gravity-mediated SUSY Breaking}
\author[andy]{Andrew Pawl  }

\address[andy]{Michigan Center for Theoretical Physics, Randall Laboratory,
University of Michigan, 500 E. University Ave., Ann Arbor, MI 48109, USA}

\begin{abstract}

To date, the properties of Q-balls arising from an Affleck-Dine
condensate in gravity-mediated SUSY breaking have been obtained
primarily through numerical simulations.  In this work, we will
derive the expected charge of the Q-balls formed in such a 
scenario through an analytical treatment.  We will also examine
the numerically observed difference between Q-ball formation in
weakly charged condensates and formation in strongly charged 
condensates.

\end{abstract}

\end{frontmatter}

\section{Introduction}

A Q-ball is a non-perturbative solution of the equation of motion
for a scalar field which is charged under a continuous U(1) symmetry
\cite{coleman1}.  The MSSM requires several such scalars, and so
it is expected that Q-balls could be formed in a supersymmetric
universe \cite{kusenko1}.  An obvious context in which to look
for Q-balls is the (baryon charged) scalar condensate that is
required for Affleck-Dine (AD) baryogenesis \cite{AD}.  
This possibility has been explored both analytically using
linear perturbation theory and numerically.  

On the analytical side, two distinct methods of formation
have been discussed in the literature.  A strongly charged condensate
might fracture into Q-balls.  This possibility is discussed in
\cite{shaposhnikov}.  It has also been noted that the characteristic
AD scalar potential in gravity mediated SUSY breaking can give rise
to negative pressure that causes perturbation growth 
\cite{EM1}.

Numerically, exhaustive simulations have been performed by Kawasaki
and Kasuya \cite{kasuya1,kasuya2}.  They observe Q-ball formation for a wide
range of initial AD field configurations, including both strongly and
weakly charged condensates (we will quantify the terms ``strongly'' and
``weakly'' later in the paper).  These simulations reveal that the mechanism
of Q-ball formation is not the same in all condensates.
For strongly charged condensates,
a proportionality between the maximum Q-ball charge
and the condensate charge density is observed.
There is, however, a distinct change in this relationship for weakly charged
condensates.  This change in scaling behavior is accompanied by (or
results from) the appearance
of Q-balls carrying a charge opposite in sign to 
that initially carried by the condensate in addition to those of
the same sign.
Thus, for weakly charged condensates, numerical simulations imply that 
Q-balls can be effectively pair-produced.

In this work, we will briefly revisit the analytical work that has been
done on Q-ball formation.  We will show that the two different approaches
to Q-ball formation can help to explain why weakly charged and strongly
charged condensates form Q-balls differently.  We will attempt to explain
the observed relationship between the charge density of the condensate
and Q-ball charge for strongly charged condensates.

\section{Review of Q-balls in the AD Scenario}
\label{sec:review}

\subsection{Flat Directions}

AD baryogenesis will require a scalar field that carries baryon number
and has a very flat potential so that it can be given a large expectation
value \cite{AD}.
There are numerous flat directions in the renormalizable MSSM without
soft SUSY breaking terms.
These 
have been cataloged \cite{martin}.  High energy 
($\ge M_{GUT}$) physics and SUSY breaking, however, ensure that no
direction will be completely flat.  In modern formulations of
the AD scenario \cite{randall1,randall2} non-renormalizable, soft-breaking,
and inflaton-coupling
terms are added to the renormalizable MSSM potential.  Such terms
lift a flat direction $\phi$ by generating a potential of the
form:
\begin{equation}
\label{eq:pot}
	U(\phi) = (m_{\phi}^{2}-c_{H} H^{2}) |\phi|^{2} +
		\frac{m_{3/2}}{M^{n-3}} (a_{m}\phi^{n} + h.c.)
		+ \frac{|\lambda|^{2}}{M^{2n-6}} |\phi|^{2n-2}
\end{equation}
where $c_{H} \sim 1$,
$|a_{m} m_{3/2}| \approx m_{\phi}
\approx$ 1 TeV (assuming gravity-mediated breaking), 
and $M$ is a large mass scale.

The phase dependent term is necessary for AD baryogenesis, as it will allow
dynamical charge creation in the condensate.  For the purpose of this
discussion, however, it is irrelevant.  We will simply assume
charge creation has occurred.
We are instead most interested in the negative mass term proportional to
the Hubble constant.  If we have a large enough $H$ during inflation,
the $\phi$ field will spontaneously move out to a finite expectation
value.  (Note that the negative sign is chosen.  It is equally likely
to be positive \cite{randall1}, but this is an uninteresting case.)

Once $H$ drops below the value
$H_{osc} = m_{\phi}c_{H}^{-1/2}$, however,  we expect the positive mass
term to begin to dominate, and $\phi$ will begin coherent oscillations
about zero.  We can find the value of $\phi$ when the oscillations
begin.  For $H \sim H_{osc}$, minimizing the potential
 (\ref{eq:pot}) yields a $\phi$ vacuum value of:
\begin{equation}
\label{eq:phiosc}
        |\phi_{osc}| \approx
	 \left(\frac{m_{\phi} M^{n-3}}{\sqrt{|\lambda|^{2}(n-1)}}
        \right)^{1/(n-2)}.
\end{equation}

This gives a characteristic initial condition for the 
scalar condensate.  Q-ball formation, however, has to wait until
the phase-dependent term has become irrelevant, otherwise charge
is not conserved.  This will require
the field value to drop by a factor of order 10 from the value
of equation (\ref{eq:phiosc}).  Examining the form of the potential
(\ref{eq:pot}) tells us that for such small $\phi$ values,
we can use the effective baryon-conserving potential:
\begin{equation}
\label{eq:bcpot}
        U(\phi) = m_{\phi}^{2}(\phi) |\phi|^{2} 
\end{equation}
where the $\phi$ dependence of $m_{\phi}$ is essential to Q-ball
stability.  We will discuss this point in the next section.

\subsection{Radiative Corrections and the Q-ball Condition}

The next step in forming Q-balls is to show that the potential 
(\ref{eq:pot}) is
compatible with the Q-ball condition \cite{coleman1}:
\begin{equation}
\label{eq:qball}
	\min_{\phi}\left[ \frac{U}{|\phi|^{2}}\right] 
	< \frac{1}{2}U''(|\phi|=0).	
\end{equation}

This condition can be realized when radiative corrections are 
taken into account \cite{EM1}.  The gauge interactions associated
with the components of the (composite) $\phi$ field
add a scale-dependence to the mass, so that:
\begin{equation}
	m_{\phi}^{2}(\phi) \approx m_{\phi}^{2}(\mu) 
\left(1+ K \log\left[\frac{
|\phi|^{2}}{\mu^{2}}\right]\right)
\end{equation}
where $K$ is negative, with a magnitude in the range
 0.01 to 0.1 \cite{EM1,EM2}. 
The negative value of $K$ is important, as it implies that
equation (\ref{eq:qball}) is satisfied.

The potential relevant for Q-ball formation, then, is:
\begin{equation}
\label{eq:qballV}
 U(|\phi|) = m_{\phi}^{2}(\mu) |\phi|^{2} \left(1+ K \log\left[\frac{
	|\phi|^{2}}{\mu^{2}}\right]\right)  
        + \frac{|\lambda|^{2}}{M^{2n-6}} |\phi|^{2n-2}.
\end{equation}

Given the form of this potential, the solution for
thin-walled Q-balls has a characteristic $\phi$ value \cite{coleman1,EM1}:
\begin{equation}
\label{eq:vevQball}
	|\phi_{0}| = \left(\sqrt{\frac{|K|}{|\lambda|^{2}(n-2)}}m_{\phi}
	M^{n-3}\right)^{1/(n-2)}.
\end{equation}
Note that this is almost the same expression as our expected
initial value of $\phi$ when the condensate begins oscillations
(equation (\ref{eq:phiosc})).  This is something of a problem
for Q-ball formation.  Recall
that Q-balls cannot form until the phase dependent term has
run its course.  We do not expect the AD field to be
large enough at that time to form thin-walled Q-balls.  

\subsection{Thick-Walled Q-balls}

Since AD baryogenesis will have to form thick-walled Q-balls
if any, we must consider the properties of such constructions.
These were examined in \cite{EM2}.
There it was shown that we can assume a Gaussian profile for
thick walled Q-balls.  
That is, we can take:
\begin{equation}
\label{eq:gaussian}
        \phi(r,t) =  \phi(r) e^{i\omega t} \equiv
        |\phi_{c}| e^{-r^{2}/R^{2}} e^{i\omega t}
\end{equation}
where $\phi_{c}$ and $R$ are constants describing the radial profile of
our Q-ball.

To see why this works, we must consider the equation of motion for the
radial part of $\phi$ \cite{coleman1}:
\begin{equation}
        \frac{\partial^{2} \phi}{\partial r^{2}} + \frac{2}{r}
        \frac{\partial \phi}{\partial r} = - \omega^{2} \phi
        + \frac{\partial U}{\partial \phi^{*}}.
\end{equation}
For a radial profile of the form (\ref{eq:gaussian}) and a
potential of the form (\ref{eq:qballV}) this yields:
\begin{equation}
        \frac{4r^{2}}{R^{4}} - \frac{6}{R^{2}} =
        - \omega^{2} + m_{\phi}^{2} - |K| m_{\phi}^{2} \left[1 + \log
        \left(\frac{\phi_{c}^{2}}{\phi_{0}^{2}}\right)\right]
        + |K| m_{\phi}^{2} \frac{2r^{2}}{R^{2}}
\end{equation}
where we have ignored the contribution of the non-renormalizable term
on the right hand side, because for $\phi_{c} < \phi_{0}$ it will be
negligible.

Comparing the sides of the above expression term by term then tells
us:
\begin{equation}
\label{eq:gausR}
        R^{2} = \frac{2}{|K| m_{\phi}^{2}}
\end{equation}
while
\begin{equation}
\label{eq:omega}
        \omega^{2} = m_{\phi}^{2} + 2|K| m_{\phi}^{2} \left[
        1+ \log\left(\frac{\phi_{0}}{\phi_{c}}\right)\right]
\end{equation}

We can also determine the charge of a Gaussian Q-ball.
The Noether current for our complex scalar is:
\begin{equation}
\label{eq:charge}
        n_{B} = \beta i (\dot{\phi}^{*} \phi - \phi^{*} \dot{\phi})
\end{equation}
where $\beta$ is the baryon charge per meson of the field $\phi$.
Now we can use equations (\ref{eq:gaussian}) and (\ref{eq:charge})
to see that the total charge 
can be expressed \cite{EM2}:
\begin{equation}
\label{eq:phiccharge}
        Q = 8\pi \beta\omega \phi_{c}^{2} \int_{0}^{\infty}
        e^{-2r^{2}/R^{2}} r^{2} dr
         = 2 \left(\frac{\pi}{2}\right)^{3/2} \beta \omega
        \phi_{c}^{2} R^{3}.
\end{equation}

\section{Strongly Charged Condensates}

\subsection{Background}

Having studied the properties of Q-balls, we now investigate
whether they are likely to be formed in an AD baryogenesis
scenario.  
For a strongly charged condensate, we anticipate that the
uniform scalar field could simply fracture into stable Q-balls.  In the
last section, we showed that the Q-balls will have a characteristic
size.  Therefore, the process of breakup
should yield Q-balls with a charge $Q$ proportional to
the initial charge density of the condensate $q_{0}$ \cite{shaposhnikov}
where $q_{0}$ in the simulations of \cite{kasuya1,kasuya2}
is defined as the charge density when 
$H \approx m_{\phi}$. 

The exact proportionality will depend on the time 
at which breakup occurs.  This is because in an expanding universe
the baryon charge density will fall with time.  For a matter-dominated
universe (Affleck-Dine baryogenesis
occurs during the epoch of inflationary reheating when
inflaton matter dominates the energy density of the universe), 
the charge density decreases as $q \propto t^{-2}
\propto H^{2}$.

\subsection{Q-ball Stability}

We first show that a strongly charged condensate is energetically
capable of breaking into stable Q-balls.
Recall that Coleman's condition for Q-ball stability is that the
energy per charge in the Q-ball be less than the energy per charge
in free particles of the scalar field in question.  The energy of
free scalars, however, is simply the mass of the field:  $m_{\phi} = U''(0)/2$.
This line of reasoning leads directly to equation (\ref{eq:qball})
\cite{coleman1}.  That equation tells us that the binding energy 
per charge for a thin-walled Q-ball will be:
\begin{equation}
	\frac{1}{\beta}\left(\frac{1}{2}U''(0)\right)^{1/2} - 
	\frac{1}{\beta}\left(\frac{U(|\phi_{c}|)}
	{\phi_{c}^{2}}\right)^{1/2}
\end{equation}
where $\phi_{c}$ is the field value within the Q-ball.  
Note that this binding energy can be maximized in 
gravity mediated scenarios at the specific value 
$\phi_{0}$ defined in equation (\ref{eq:vevQball}).

A problem arises when we try to calculate the value of the binding
energy, however.  It is a well-known difficulty (see e.g. \cite{coleman2})
that the running mass of our scalar field has an infrared divergence
associated with $\phi = 0$.  Even worse, however, the one-loop running
becomes untrustworthy for large values of the log term, since the next
loop order will bring in higher powers of the log.  
This implies that we cannot actually
calculate the mass of a free $\phi$ meson by one-loop running from
the value of the mass at large $\phi$ and, more importantly, the qualitative
behavior becomes suspect at large values of the log term.  

This could be a very important issue for thick-walled Q-balls, which
necessitate $\phi_{c} << \phi_{0}$ so that we are already stretching
the one-loop running to the limits of applicability. 
If we want to have a truer idea of the Q-ball
binding energy, we should try to make use of the renormalization
group, or at the very least move to two-loop running.  
This has not been attempted in the numerical simulations to date,
however.  Therefore, let us consider the consequences of assuming
simple one-loop running holds over a large range of $\phi$.

With this assumption, we see that although $\phi_{0}$ maximizes the binding
energy per charge,
the binding energy will remain positive (formally infinite) for any value 
$\phi < \phi_{0}$.  This is very important for us, since we have
already shown that the AD scenario will only generate thick-walled
Q-balls.  We can now see that they are potentially stable.
Unfortunately, our discussion so far is not technically valid for thick-walled
Q-balls since they will require the addition of a gradient term to
the expression for the Q-ball energy.  

The energy of a complex
scalar field is given by \cite{coleman1}:
\begin{equation}
\label{eq:energy}
  E = \int d^{3}x \left[\dot{\phi}\dot{\phi}^{*} + \vec{\nabla}\phi\cdot
\vec{\nabla}\phi^{*} +U(|\phi|)\right].
\end{equation}
Note that if we do have gradient energy, we can treat it as an effective
contribution to the scalar potential.  That is, we can define an
effective potential for thick-walled Q-balls:
\begin{equation}
	\tilde{U} = U + \vec{\nabla}\phi\cdot
	\vec{\nabla}\phi^{*}.
\end{equation}

We can estimate the contribution of the gradient term to $\tilde{U}$
by using the Gaussian approximation.  Recall that thick-walled
Q-balls have the profile:
\begin{equation}
        \phi(r,t) =  
        |\phi_{c}| e^{-|K|m_{\phi}^{2}r^{2}/2} e^{i\omega t}
\end{equation}
which means:
\begin{equation}
	\vec{\nabla}\phi\cdot
        \vec{\nabla}\phi^{*}= |K|^{2}m_{\phi}^{4}r^{2} 
	e^{-|K|m_{\phi}^{2}r^{2}} |\phi_{c}|^{2}.
\end{equation}	

Looking at this equation, we see that the gradient term will
make a contribution to the potential which is at maximum
of order $|K|m_{\phi}^{2}|\phi_{c}|^{2}$.  Since the 
energy per charge has the form:
\begin{equation}
	\frac{E}{Q} = \frac{1}{\beta}\sqrt{\frac{U(|\phi_{c}|)}{|\phi_{c}|^{2}}}
\end{equation}
we expect the gradient energy to yield a
contribution of order $|K|^{1/2}m_{\phi}$.  We would like to
know if this is significant on the scale of the expected
binding energy per charge.

Even though the binding energy is formally infinite, we can
still compare the characteristic scale of the binding energy
with the characteristic scale of the gradient energy.  
In the thick-walled case, we expect that the 
potential will be dominated by the mass term.  Thus, we can estimate
the binding energy per baryon as:
\begin{equation}
\label{eq:approxbe}
	\frac{1}{\beta}\sqrt{m^{2}_{\phi}(0)}
	- \frac{1}{\beta}\sqrt{m^{2}_{\phi}(\phi_{c})}
	\approx \frac{1}{\beta} |K|^{1/2}m_{\phi} \log(\phi_{c}/0).
\end{equation}
This equation shows that although the binding energy is formally infinite,
it has the characteristic magnitude $|K|^{1/2}m_{\phi}$ (we will
assume that the log running is somehow regulated by higher-order
effects).  This implies that we can expect a binding energy per
charge of at least a few times $|K|^{1/2} m_{\phi}$, and
so the gradient energy of thick-walled Q-balls is not too large
to admit bound Q-balls.
These conclusions are borne out in numerical simulations.  In
\cite{kasuya1,kasuya2} Q-balls are shown to form over a wide
range of initial $\phi$ values (many decades). 

\subsection{Partially Charged Q-balls}

So far we have only considered Q-balls which follow a circular
path in phase space ($\omega$ is constant).  As we shall see,
this corresponds to a scalar field that is completely charge
asymmetric (all baryons or all anti-baryons).  This is not
a necessary condition for a strongly charged condensate, however.
In \cite{kasuya1} it is seen that even condensates with considerable
ellipticity in their phase space rotation will form Q-balls that
obey the same scaling relations as for completely charged condensates.
This implies that elliptically rotating condensates can fracture
into bound Q-balls as well.  We would like to examine the energetics
of this process.

The first step is to parameterize
an elliptical
condensate.  Recall our effective potential after the phase dependent terms
have become unimportant is:
\begin{equation}
	U(|\phi) = m_{\phi}|\phi|^{2}
	\left(1-|K|\log\left(\frac{\phi^{2}}{\phi_{1}^{2}}\right)\right)
\end{equation}
where $\phi_{1}$ is basically arbitrary.  As long as $\phi$ does not 
change more than an order of magnitude or so, the log corrections are
essentially unimportant assuming small $|K|$ (as we shall see, this
is true for a condensate with significant charge).  Therefore we effectively
have a $|\phi|^{2}$ potential.  Such a potential allows for closed
orbits in the complex $\phi$ plane \cite{goldstein}.  We can parameterize
such an orbit in the form:
\begin{equation}
\label{eq:param}
	\phi = A \cos(m_{\phi} t) + iB\sin(m_{\phi} t).
\end{equation}

Note that in keeping with our orbit analogy 
the analog of angular momentum in this case is $\dot{\theta}_{\phi}|\phi|^{2}$
which, by the Noether current, is equal to $n_{B}/2\beta$ where $n_{B}$ is the
number density of excess baryons.
Conservation of angular momentum
is guaranteed by the central force character of our potential (even with
logarithmic correction).  
Plugging our parameterization into the Noether current yields:
\begin{equation}
\label{eq:ellipchg}
 	n_{B} = 2\beta m_{\phi}AB.
\end{equation}

We can now compare
this number to the total number of $\phi$ particles that our condensate
contains.  This is approximated by taking the total energy
density
in the condensate, as given by equation (\ref{eq:energy}),
and dividing by $m_{\phi}$:
\begin{equation}
\label{eq:smchgn}
	n_{\phi} \approx m_{\phi} A^{2} + m_{\phi} B^{2}.
\end{equation}
It is now apparent that if we have a circular ``orbit'' the condensate is
entirely composed of baryons (or antibaryons).  For elliptical orbits,
higher eccentricity means less net baryon fraction.  

Now let us evaluate the energy per charge of an elliptical Q-ball.  
We can make the analysis easier by picking a particular point along
the trajectory.  The simplest to use is when $\phi=A$.
Using our small $\phi$ approximation for the potential and ignoring
gradient terms for simplicity, we obtain:
\begin{equation}
\label{eq:smchgen}
	E = (m_{\phi}^{2}B^{2} + m_{\phi}^{2} A^{2})V
\end{equation}
where $V$ is the volume occupied by the Q-ball.  The charge of the Q-ball
will be $n_{B} V$ where $n_{B}$ is given in equation (\ref{eq:ellipchg}).
Thus, we have an energy per charge:
\begin{equation}
\label{eq:problem}
	\frac{E}{Q}
	 = \frac{1}{2\beta}\left(m_{\phi} \frac{B}{A}+
	m_{\phi}\frac{A}{B}\right).
\end{equation}
Note that throughout this analysis we have ignored the running of the mass
with $\phi$.  Now we can see that this was justified.  The characteristic
contribution of the ellipticity to the energy per charge is of order $m_{\phi}$,
while the binding energy is characteristically of order $|K|^{1/2}m_{\phi}$.

This is a serious concern.  By comparing equation (\ref{eq:problem}) to
the result of equation (\ref{eq:approxbe}) we can
see that once $A$ is a few times greater than $B$ it
would require the one-loop mass running to be valid over several orders of
magnitude to yield a bound Q-ball.  In contrast, numerical simulations 
show that elliptical condensates can certainly form Q-balls up to ratios
$A/B \sim 10$, and possibly beyond \cite{kasuya1}.  

We can reconcile this apparent discrepancy by noting that the cause of the
instability of Q-balls with a large energy per net baryon is quantum mechanical
decay.  Such decay is not treated in numerical simulations of the classical
field.  Thus, our estimates would lead us to conclude that unless some way
of damping out the ellipticity of a partially-charged AD condensate is 
present, we must question whether such condensates will truly form Q-balls.

One possibility for such damping is scalar self-interaction in a nonhomogeneous
condensate.  It is possible that this has already been observed in the
numerical work of \cite{kasuya1,kasuya2} and simply not published.  Perhaps
a more interesting possibility, however, is that in this case the problem
is also the cure.

A rotating scalar condensate need not decay only to free scalar particles.
We anticipate couplings of the AD field to many other particle species.
In this case, we would anticipate that a dense condensate could experience
annihilations that remove the charge-neutral part while leaving behind the
baryon asymmetry.  This would effectively remove the excess energy per 
baryon.

  We have calculated the annihilations that would be
mediated by 
an interaction term of the form $|\phi|^{2}|\chi|^{2}$ where $\chi$ is another
scalar field.  Such an interaction would generically be present for a flat
direction composed of squarks.  Our calculations show that the annihilations
mediated by this interaction would reduce the ellipticity of the condensate
dramatically on a time scale that is considerably less than the time scale
for Q-ball formation \cite{apawl}.

\subsection{Breakup}

So far in this section, we have summarized the evidence that a uniformly
charged AD condensate is
in a state that meets the criteria to be considered a bound 
Q-ball immediately after beginning coherent oscillations 
about $\phi = 0$.   
On the strength of these arguments, we expect that
by the time the universe is of the
age $t \sim \mathcal{O}(1) \times m_{\phi}^{-1}$ the condensate
is energetically ready to form Q-balls.  Numerical simulations, however,
show that the time of Q-ball formation is more nearly of the order
$t \sim \mathcal{O}(1000) \times m_{\phi}^{-1}$ for
$|K| = 0.01$ \cite{kasuya1}.

This discrepancy is a result of the finite age of the universe (or,
equivalently, the expansion of the universe).  In a matter dominated
universe, the maximum radius of a causally connected patch is
\begin{equation}
	r_{cc} = 3t = 2H^{-1}.
\end{equation}
Thus, even if the scalar field would energetically prefer to form
Q-balls, the Q-ball sized patches cannot form a coherent object
until $r_{cc} \approx |K|^{-1/2}m_{\phi}^{-1}$.  

For $|K| = 0.01$ Q-ball formation would then be causally allowed
by $t \sim \mathcal{O}(10) \times m_{\phi}^{-1}$.  Thus, we still
have a large factor to make up.  As we shall see, this 
is because the forces responsible for Q-ball formation 
propagate at a characteristic sound speed rather than the speed of light.

Q-ball formation is the result of interactions in the fluid of
squarks represented by the scalar condensate.  Such fluid interactions
will have a sound speed associated with them.  The sound speed
in a completely charge-asymmetric scalar condensate is given
in \cite{coleman1} as:
\begin{equation}
\label{eq:vsound}
        v_{s}^{2} = \frac{U''(|\phi|)|\phi|-U'(|\phi|)}
                {U''(|\phi|)|\phi|+3U'(|\phi|)}.
\end{equation}
Evaluating this using the effective potential of equation (\ref{eq:bcpot})
gives the result that $v_{s}^{2} \approx K/2$.  It is important
to note that because $K$ is negative, $v_{s}$ is technically imaginary. 
As we shall see, this will lead to the amplification of small
perturbations.

To see this amplification, consider the plane-wave solution 
to the wave equation:
\begin{equation}
	\psi(\vec{r},t) = \psi_{0}\exp\left(i\vec{k}\cdot \vec{r}
	\pm i \omega_{s} t\right)
\end{equation}
where for sound waves $(\omega_{s}/|k|)^{2} = v_{s}^{2}$.  

Now, consider
the situation when $v_{s}^{2}$ is negative.  We can construct a
very similar solution, except that now 
we must take $\omega_{s}$ to be imaginary.  
Thus, we anticipate solutions of
the form:
\begin{equation}
	\psi(\vec{r},t) = \psi_{0}\exp\left(i\vec{k}\cdot \vec{r}\right)
        \exp\left(\pm |\omega_{s}| t\right).
\end{equation}

This solution yields growing modes which increase exponentially
with e-folding time equal to:
\begin{equation}
	t_{e-fold} = |\omega_{s}|^{-1} = \frac{1}{|v_{s}| |k|}.
\end{equation}
Since a sound wave is constructed of small density fluctuations in
the medium, this tells us that such fluctuations will grow exponentially.
Further, since the density of the medium is proportional to $|\phi|^{2}$,
we anticipate that small fluctuations in the field behave as $\delta \rho \sim 
|\phi| \delta \phi$, so that $\phi$ fluctuations will grow on basically
the same time scale as density perturbations.

Consider what this means for our condensate.  We expect to form Q-balls
with characteristic diameter $2^{3/2}|K|^{-1/2}m_{\phi}^{-1}$.  Thus, to
decide how fast perturbations on this scale will grow, we can use this
diameter as half the wavelength in determining $|k|$ in the above equation.
We have already calculated $v_{s}$.  Combining these results with 
the results above gives:
\begin{equation}
	t_{e-fold} \approx \frac{2\sqrt{2}}{\pi |K| m_{\phi}} \approx \frac{1}
	{|K|m_{\phi}}.
\end{equation}
For $|K|=0.01$, this
evaluates to a characteristic time scale for the growth of Q-ball-sized
perturbations:
\begin{equation}
	t_{e-fold} \approx 100 m_{\phi}^{-1}.
\end{equation}
Further, since the numerical simulations of interest start off with
perturbations suppressed by a factor of $10^{7}$ \cite{kasuya1},
we anticipate Q-ball formation will require about 16 e-folding times.
This yields exactly the relation found in the numerical simulations:
\begin{equation}
	t_{Q-ball} \approx 16 t_{e-fold} \approx 3000 m_{\phi}^{-1}.
\end{equation}
Calculating the total charge contained in the Q-ball sized perturbation
at this time, we find that:
\begin{equation}
	Q \approx \frac{4\pi}{3}R_{Q}^{3} q(0)t_{Q-ball}^{-2}m_{\phi}^{-2}
	  \approx 0.0057 m_{\phi}^{-3}q(0)
\end{equation}
which compares very well with the results obtained on three-dimensional
lattices in \cite{kasuya2}.

Before we leave this section, we should comment that this analysis is
only valid for imaginary values of the sound speed in Q-matter.  Looking
back at equation (\ref{eq:vsound}), this means that Q-ball formation
will require:
\begin{equation}
	U''(|\phi|)-U'(|\phi|)/|\phi| < 0.
\end{equation}
This condition can be restated using the equation of motion for $|\phi|$.
From this, we can see that the phase rotation rate $\omega$ for a
completely charge-asymmetric condensate is given
by \cite{coleman1}:
\begin{equation}
	\omega^{2} =U'(|\phi|)/|\phi|.
\end{equation}
Thus, we can rewrite our condition for Q-balls:
\begin{equation}
	U''(|\phi|)-\omega^{2} < 0
\end{equation}
which is precisely the condition for Q-ball formation in a completely
charge-asymmetric condensate arrived at in
\cite{shaposhnikov} using the equations of motion for $|\phi|$
and $\omega$.  

\section{Weakly Charged Condensates}

\subsection{Background}

For condensates with little or no net charge, we expect that breakup
into Q-balls must follow a different path than for strongly charged
condensates.  The only way to form Q-balls is to somehow separate the
neutral condensate into positive and negatively charged regions.  
This is observed to happen in numerical simulations 
\cite{kasuya1}.  We wish to try to understand this process analytically.

\subsection{Negative Pressure and Growth of Perturbations}

In one of the early papers on Q-balls in gravity-mediated AD scenarios,
it was assumed that Q-ball formation was due to negative pressure
in the AD condensate \cite{EM1}.  For the case of a real (non-complex)
scalar field, negative pressure condensates result
from the negative value of $K$ in the one-loop corrected scalar potential
(see equation (\ref{eq:qballV})) \cite{turner,mcdonald}.  We wish to
explicitly consider the pressure arising for a charged (complex) scalar.

It is important to note that in this section we will have to be 
careful to retain terms of order $|K|$ in the expressions we are
about to derive.  These small corrections will be responsible for
any pressure that we find.  The first such correction to take note
of is that when we parameterize our rotating condensate in the
form of an ellipse, we must use the more general form:
\begin{equation}
\label{eq:paramom}
        \phi = A \cos(\omega t) + iB\sin(\omega t)
\end{equation}
where to first order in $|K|$, $\omega$ is given by:
\begin{equation}
\label{eq:ompert}
	\omega = m_{\phi}\sqrt{1-|K|}.
\end{equation}
(This result follows from the equation of motion.)  Note that
for a treatment of ``weakly charged'' condensates we are interested
in $A/B > 10$.

With this form in hand, we can move on to explicitly calculate the
pressure.
In \cite{turner} it was shown that for a rapidly (relative to the 
Hubble time) oscillating condensate, the average pressure can
be written in terms of the energy density through the use of the
parameter:
\begin{equation}
	\gamma = \frac{1}{T}\int_{t}^{t+T} \frac{p+\rho}{\rho} dt
\end{equation}
where $p$ is the pressure, $\rho$ is the energy density, and $T$ is
the period of the oscillation.  In terms of $\gamma$, we would have
the relationship between average pressure and energy density:
\begin{equation}
	\bar{p} = (\gamma-1)\rho.
\end{equation}

To evaluate this expression for our charged condensate we must use
the energy-momentum tensor for a complex scalar:
\begin{equation}	
	T^{\mu}_{\;\;\nu} = \frac{\partial \mathcal{L}}{\partial
	\partial_{\mu}\phi} \partial_{\nu} \phi +
	 \frac{\partial \mathcal{L}}{\partial
        \partial_{\mu}\phi^{*}} \partial_{\nu} \phi^{*} - \mathcal{L}
	\delta^{\mu}_{\;\;\nu}
\end{equation}
which tells us:
\begin{equation}
\label{eq:dotsolve}
	\rho = \dot{\phi}\dot{\phi}^{*} + U
\end{equation}
and:
\begin{equation}
	p = \dot{\phi}\dot{\phi}^{*} - U.
\end{equation}

These, along with the generic elliptical parameterization of
equation (\ref{eq:paramom}), yield the expressions:
\begin{equation}
	\rho + p = 2 \dot{\phi}\dot{\phi}^{*} = 2\left(\rho-U\right)
\end{equation}
and:
\begin{equation}
\label{eq:realenden}
	\rho = m_{\phi}^{2}\left(A^{2}+B^{2}\right)-|K|m_{\phi}^{2}B^{2}.
\end{equation}

\begin{figure}
\begin{center}
\psfig{file=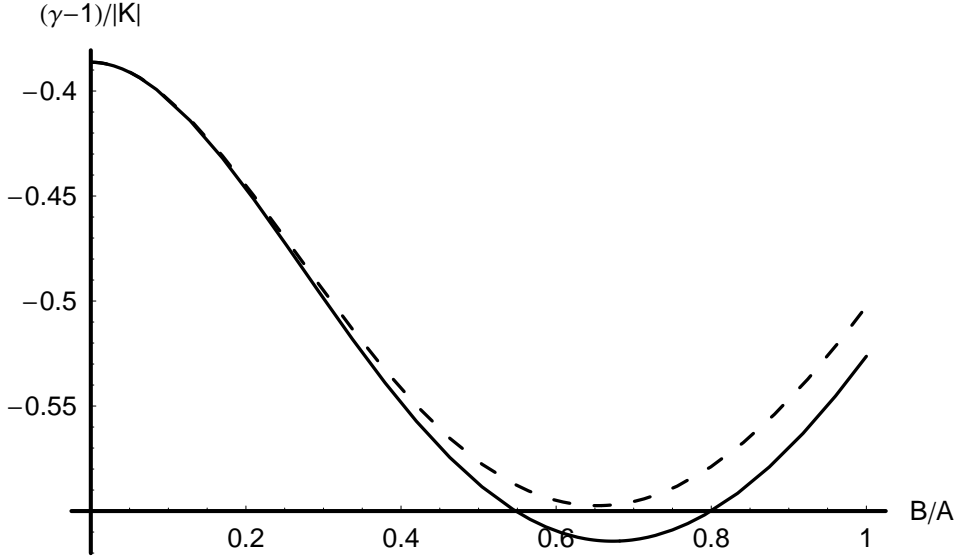,width=5.0in}
\caption{Plot of $p/(|K|\rho)$ for condensates with varying ellipticities. 
The solid line is $K = -0.1$, while the dotted line is $K=-0.01$.}
\label{fig:gam}
\end{center}
\end{figure}

Now, assuming that $\rho$ changes very little over the course of
one oscillation, we can write:
\begin{equation}
	\gamma = \frac{\omega}{\pi} \int_{0}^{2\pi/\omega} 
	\left(1-\frac{U}
	{m_{\phi}^{2}A^{2} + \omega^{2}B^{2}}\right) dt.
\end{equation}
This expression can be numerically integrated.  Doing so for
various values of $B/A$ and $K$ yields the curve shown in Figure \ref{fig:gam}.
The feature of current interest is the intercept at
$B = 0$, which is consistently equal to $-0.39|K|$. 

With this result we see that the analysis of \cite{EM1} is essentially
correct.  The pressure of a weakly charged AD condensate is expected to be
negative,
with $\gamma-1 \approx K/2$.  Note, however, that we have made something
of an approximation in our parameterization of the AD condensate as
a perfect ellipse.
A study of ``true'' AD condensates obtained via
numerical integration of the classical equations of motion shows that
the numerical treatment for the idealized elliptical trajectory
is accurate to within a factor
of $2$.  Therefore, we have settled on $K/2$ as a reasonable first
approximation to the pressure.  Note also that this is the result
obtained using the 
polynomial approximation of \cite{mcdonald}.

We can now follow \cite{EM1} to see that perturbations in even
a perfectly charge-symmetric AD condensate will grow exponentially.  In
fact, since the factor $\gamma -1$ is identical to the sound speed $v_{s}$ 
found for a strongly charged condensate, the exponential growth will
occur with the same timescale as the growth of perturbations in a 
strongly charge-asymmetric condensate.  

\subsection{Charge Separation}

The negative pressure of the condensate can serve to give us growing
Q-ball sized perturbations.  The next question, however, is how these
regions can acquire a charge asymmetry.  The dynamics of such a process
are beyond a simple perturbative analysis.  We can, however, do some
back of the envelope calculations to guess at the charges we could expect.

Consider two overdense Q-ball-sized perturbations that have evolved 
``next to'' one another.  We will consider the epoch when these perturbations
have just gone nonlinear, so $\delta \equiv \delta \rho/ \bar{\rho} \approx 1$
where $\bar{\rho}$ is the average energy density in the scalar field and
$\delta \rho$ is the peak energy density in the perturbation.  The perturbations
will be separated by an underdense region
of approximately Q-ball size.  

We must assume that they have interacted
in such a way as to impart on one another a very weak rotation.  By 
conservation of baryon number we know one will have a rotation 
corresponding to positive baryon asymmetry and the other will carry
negative baryon asymmetry.  We now wish to show that this weak rotation
is enough to cause the positive perturbation to preferentially accrete
baryons as it grows, while the negative perturbation will preferentially
accrete anti-baryons.  Thus the charge asymmetries will be enhanced, eventually
yielding a Q-ball/anti-Q-ball pair.

Consider a tiny box (side length $L$) filled 
with baryons that lies on the line joining
the two perturbations.  Imagine that it is closer to the center of the
negative perturbation.  We can calculate work required to remove this
box from the negative perturbation.  The work will be:
\begin{equation}
	W_{rm} = \int_{r_{0}}^{\infty} \Delta p L^{2} dr
\end{equation}
where $r_{0}$ is the distance between the center of the negative perturbation
and the small box to begin with, and 
 $\Delta p$ is the pressure difference between the leading edge
and the trailing edge of the box, given approximately by 
\begin{equation}
	\Delta p \approx - \frac{dp}{dr} L.
\end{equation}
So, using the results of the previous section:
\begin{equation}
	W_{rm} = \int_{r_{0}}^{\infty} L^{3} \frac{|K|}{2} \frac{d\rho}{dr} 
	= L^{3} \frac{|K|}{2} \rho(r_{0}).
\end{equation}

We want to compare this work with the energy difference that we would expect
to see if the box of baryons left the negative perturbation and joined
the positive perturbation.  
We can estimate this by considering the perturbations to be highly elliptical
Q-balls.  Thus, we expect that the perturbations contain net baryon number
$N_{B}$
equal to (see equation
(\ref{eq:smchgn})):
\begin{equation}
	N_{B} = \pm 2 \beta m_{\phi} AB V_{Q}
\end{equation}
where the minus denotes the negative perturbation and $V_{Q}$ is the
volume of the perturbation.
Similarly, each perturbation has an energy (see equation (\ref{eq:smchgen})):
\begin{equation}
	E = (m_{\phi}^{2} A^{2} + m_{\phi}^{2} B^{2})V_{Q}
\end{equation}
where in each of these formulas, we expect $A \gg B$.

Adding the box of baryons to either perturbation implies a change in both
the total baryon asymmetry and the energy.
For the net baryon number, we have:
\begin{equation}
	\delta N_{B} = \beta \frac{\rho_{b}}{m_{\phi}} L^{3}
	= \pm 2 \beta m_{\phi} \left(A \delta B + B \delta A\right)
	 V_{Q} 
\end{equation}
where $\rho_{b}$ is the energy density within the small box,
and we have differentiated
$N_{B}$ given above.
The energy relation is:
\begin{equation}
	\delta E = \rho_{b}L^{3} = 2 m_{\phi}^{2} \left(A \delta A +
	B \delta B \right)
	V_{Q}.
\end{equation}

We can solve these equations simultaneously to obtain:
\begin{equation}
	\delta A_{+} = \delta B_{+}
	 = \frac{\rho_{b} L^{3}}{2 m_{\phi}^{2} V_{Q}}
	\frac{A-B}{A^{2}-B^{2}}
\end{equation}
for the positive perturbation, while for the negative perturbation
we would have:
\begin{equation}
	\delta A_{-} = - \delta B_{-} = 
	\frac{\rho_{b} L^{3}}{2 m_{\phi}^{2} V_{Q}}
        \frac{A+B}{A^{2}-B^{2}}.
\end{equation}

We have solved for these by explicitly assuming the energy contribution
of the box of baryons is the same for each perturbation.  To zeroth order
in $K$ we expect this to be the case, as there is no attractive force
present.  Once we ``turn on'' $K$, however, we do expect to see an attraction
between baryons.  We can estimate this attraction by using our zeroth
order solutions for $\delta A$ and $\delta B$ in the first order expression
for energy.  From equation (\ref{eq:realenden}) we can see:
\begin{equation}
	E_{unlike} - E_{like} = 2 |K|\rho_{b} L^{3} \frac{BA}{A^{2}-B^{2}} 
	\approx 2 |K| \rho_{b} L^{3} \frac{B}{A}.
\end{equation}
where ``like'' denotes the energy of a positive perturbation with the positive
box of baryons or of a negative perturbation with a negative box, and ``unlike''
represents the other two possibilities.
Thus we see that at first order in $K$ the energy of the box is lower 
when it is within the positive Q-ball than when it is within the negative
Q-ball.  This energy difference can provide the work needed to separate the
negative charges from the growing positive perturbation and vice-versa.

\subsection{Q-ball Charge}

We will now make a very rough estimate of the amount of charge separation
that will occur in the formation of a perturbation from a weakly charged
condensate.  

In the simplest picture, we are forming spherical perturbations with
radius $R$ (approximately $|K|^{-1/2}m_{\phi}^{-1}$) which are separated
from one another by voids of about $2R$ in width.  Thus, each perturbation
has a sphere of influence that extends a radius $2R$ from its center. 
Theoretically it will gather up all the matter in this sphere to create
the final perturbation.  

We will now assume that we have a
pair of oppositely charged perturbations next to one another.  Now
in the region between these, the baryons will be preferentially attracted
to the positive perturbation, while the antibaryons will be preferentially
attracted to the negative one.  Thus, we will somewhat arbitrarily define
an extended sphere of influence for these charged Q-balls of radius $3R$.
In the overlap region of these extended spheres, we will assume that 
the charges separate completely and collect in their energetically 
preferred Q-ball.

To determine the final charge, then, we are interested
in the ratio of this overlap volume to that of the complete $2R$ sphere
of influence.  A quick calculation shows that the ratio in question is
approximately 1/10.  Thus, we expect a final charge fraction in the 
Q-ball of 1/10 the value that a Q-ball formed in a completely charge asymmetric
condensate of the same initial $|\phi|$ value would have.  This is in
reasonable agreement with numerical simulation \cite{kasuya1,kasuya2}.  

\section{Summary}

        We anticipate that strongly charged condensates can easily
break up into bound Q-balls.  It is clear from a simple analysis, however,
that any significant ellipticity in the phase rotation of the
condensate might compromise this binding by
making the energy per charge larger than that in a free scalar plasma.
This concern leads us to believe that damping of this ellipticity is
important for Q-ball formation.  This has been considered in
a separate work \cite{apawl}.  (Note added:  After publication
of this paper, I was informed that the delayed decay of Affleck-Dine 
condensates has also been considered in \cite{postma}.)

        Assuming that strongly charged condensates can reach a nearly
circular rotational state, we recover the known result \cite{shaposhnikov}
that breakup will occur.  By considering the sound speed in the
scalar condensate, we have shown that the time scales for breakup observed
in numerical simulations \cite{kasuya1,kasuya2} are reproduced in
a simple analytical treatment.  Given the results of \cite{kasuya1,kasuya2}
we expect these ``strongly charged'' results to be useful for
condensates with $A/B$ as high as $\mathcal{O}(10)$.  

	For condensates with even larger $A/B$ ratios, we have shown that
the negative pressure of the condensate \cite{EM1,mcdonald} can still
allow for the growth of seed perturbations.  This, coupled with the weak
attractive force between baryons can lead to Q-ball/anti-Q-ball pairs.

\end{document}